\documentclass[aps,prd,twocolumn,superscriptaddress]{revtex4}
\usepackage{epsfig,epsf}
\usepackage{amsmath}
\usepackage{amsthm}
\usepackage{amsfonts}
\usepackage{amssymb}
\usepackage{dsfont}
\usepackage{multirow}
\usepackage{appendix}
\usepackage{slashed}
\usepackage[active]{srcltx}
\usepackage{psfrag}

\begin{document}

\title{Parameters of the tensor tetraquark $bb\overline{c}\overline{c}$}
\date{\today}
\author{S.~S.~Agaev}
\affiliation{Institute for Physical Problems, Baku State University, Az--1148 Baku,
Azerbaijan}
\author{K.~Azizi}
\thanks{Corresponding Author}
\affiliation{Department of Physics, University of Tehran, North Karegar Avenue, Tehran
14395-547, Iran}
\affiliation{Department of Physics, Do\v{g}u\c{s} University, Dudullu-\"{U}mraniye, 34775
Istanbul, T\"{u}rkiye}
\affiliation{Department of Physics and Technical Sciences, Western Caspian University,  Baku, AZ 1001, Azerbaijan}
\author{H.~Sundu}
\affiliation{Department of Physics Engineering, Istanbul Medeniyet University, 34700
Istanbul, T\"{u}rkiye}

\begin{abstract}
The mass and width of the tensor tetraquark $T=bb\overline{c}\overline{c}$
with spin-parity $J^{\mathrm{P}}=2^{+}$ are calculated in the context of the
QCD sum rule method. The tetraquark $T$ is modeled as a diquark-antidiquark
state built of components $b^{T}C\gamma _{\mu }b$ and $\overline{c}\gamma
_{\nu }C\overline{c}^{T}$ with $C$ being the charge conjugation matrix. The
mass $m=(12.795\pm 0.095)~\mathrm{GeV}$ of the exotic tensor meson $T$ is
found by means of the two-point sum rule approach. Its full width $\Gamma$
is evaluated by considering processes $T \to B_{c}^{-}B_{c}^{-}$, $
B_{c}^{-}B_{c}^{\ast -}$, and $B_{c}^{\ast -}B_{c}^{\ast -}$. Partial widths
of these decays are computed by means of the three-point sum rule approach
which is used to determine the strong couplings at relevant
tetraquark-meson-meson vertices. Predictions obtained for the width $\Gamma
_{\mathrm{T}}=55.5_{-9.9}^{+10.6}~\mathrm{MeV}$, as well as the mass of the
tetraquark $T $ can be useful in investigations of fully heavy four-quark
mesons.
\end{abstract}

\maketitle


\section{Introduction}

\label{sec:Intro}

Fully heavy exotic mesons containing four $b$ and/or $c$ quarks recently
became objects of intensive studies. This interest is triggered by
impressive experimental achievements of a last few years when different
experimental groups reported about observation of such structures. Thus, the
LHCb, ATLAS, and CMS Collaborations discovered four new $X$ resonances
presumably composed of four $c$ ($\overline{c}$) quarks in the mass range $%
6.2-7.3~\mathrm{GeV}$ \cite{LHCb:2020bwg,Bouhova-Thacker:2022vnt,CMS:2023owd}%
. They were seen in the mass distributions of $J/\psi J/\psi $ and $J/\psi
\psi ^{\prime }$ mesons and provided valuable information to understand
internal organizations of multiquark systems.

Experimental data are also crucial to test different models and schemes
suggested to compute parameters and explain features of these resonances.
Theoretical models treat $X$ states mainly as diquark-antidiquark or
hadronic molecule-type compounds \cite%
{Zhang:2020xtb,Albuquerque:2020hio,Yang:2020wkh,Becchi:2020mjz,Becchi:2020uvq, Wang:2022xja,Faustov:2022mvs,Niu:2022vqp,Dong:2022sef,Yu:2022lak, Kuang:2023vac}%
. In the context of an alternative point of view one considers them as
coupled-channel effects, which manifest themselves in a form of known
resonances \cite{Dong:2020nwy,Liang:2021fzr}.

The $X$ structures were also modeled and studied in a rather detailed form
in our publications \cite%
{Agaev:2023wua,Agaev:2023ruu,Agaev:2023gaq,Agaev:2023rpj}. In these papers
we computed not only their masses but estimated also full widths of these
tetraquarks. Some of the $X$ resonances were interpreted as ground-level or
excited diquark-antidiquark states, whereas for others preferable models are
admixtures of diquark-antidiquark and molecule-type structures. It should be
noted that theoretical investigations of fully heavy tetraquarks cover a
considerably wide time span. Here, we have mentioned publications in which
authors analyzed LHCb-ATLAS-CMS data: Relatively complete list of previous
articles can be found in Ref.\ \cite{Agaev:2023wua}.

The exotic mesons $bb\overline{c}\overline{c}$ form a new cluster of
all-heavy four-quark mesons. Some of such four-quark mesons may have a mass
below relevant $B_{c}B_{c}^{(\ast )}$ thresholds and be a strong-interaction
stable particle, therefore attracts an interest of researchers. The reason
is that, the quark content $bb\overline{c}\overline{c}$ forbid their strong
decays through $b\overline{b}$ or $c\overline{c}$ annihilations which makes
them promising candidates to stable tetraquarks. This feature distinguish $bb%
\overline{c}\overline{c}$ and $bb\overline{b}\overline{b}$ (or $cc\overline{c%
}\overline{c}$) tetraquarks because latter are unstable even residing below
two-particle bottomonia (or charmonia) limits \cite%
{Becchi:2020mjz,Becchi:2020uvq,Agaev:2023ara}.

During last years the structures $bb\overline{c}\overline{c}$ were
investigated using numerous methods \cite%
{Wu:2016vtq,Li:2019uch,Wang:2019rdo,Liu:2019zuc,Chen:2019vrj,Wang:2021taf,Mutuk:2022nkw,Galkin:2023wox}%
. As a rule, the masses of these tetraquarks were fixed above corresponding
two-meson thresholds. But in accordance with Ref.\ \cite{Wang:2021taf},
scalar, axial-vector and tensor tetraquarks $cc\overline{b}\overline{b}$
with special internal organizations are stable against strong decays.
Consequently, they can transform to ordinary particles only through
electroweak decays.

In our articles \cite{Agaev:2023tzi,Agaev:2024pej}, we explored the scalar
and axial-vector tetraquarks $bb\overline{c}\overline{c}$ by means of the
sum rule approach. The scalar exotic meson $X_{\mathrm{1}}$ composed of an
axial-vector diquark and antidiquark, and $X_{\mathrm{2}}$ built of
pseudoscalar diquarks were studied in Ref.\ \cite{Agaev:2023tzi}. The masses
of these particles are equal to $(12715\pm 80)~\mathrm{MeV}$ and $(13370\pm
95)~\mathrm{MeV}$ which exceed kinematical limits for production of $%
B_{c}^{-}B_{c}^{-}$ mesons. We evaluated the full widths $\Gamma _{\mathrm{1}%
(\mathrm{2})}$ of the tetraquarks $X_{\mathrm{1}}$ and $X_{\mathrm{2}}$ by
studying their decay modes $X_{\mathrm{1}}\rightarrow B_{c}^{-}B_{c}^{-}$
and\ $B_{c}^{\ast -}B_{c}^{\ast -}$, and $X_{\mathrm{2}}\rightarrow
B_{c}^{-}B_{c}^{-}$,\ $B_{c}^{-}B_{c}^{-}(2S)$, and$\ B_{c}^{\ast
-}B_{c}^{\ast -}$, respectively. Results obtained for $\Gamma _{\mathrm{1}%
}=(63\pm 12)~\mathrm{MeV}$ and $\Gamma _{\mathrm{2}}=(79\pm 14)~\mathrm{MeV}$
confirm that they are tetraquarks with relatively modest widths. The
quantitatively similar predictions were found in Ref.\ \cite{Agaev:2024pej}
for axial-vector states $T_{\mathrm{1}}$ and $T_{\mathrm{2}}$: Full widths
of these particles amount to $(44.3\pm 8.8)~\mathrm{MeV}$ and $(82.5\pm
13.7)~\mathrm{MeV}$,respectively. It turned our, that all tetraquarks
analyzed in these two papers are strong-interaction unstable structures.

In present work, we study the exotic tensor meson $T=bb\overline{c}\overline{%
c}$ with quantum numbers $J^{\mathrm{P}}=2^{+}$. We model $T$ as a
tetraquark made of diquarks $b^{T}C\gamma _{\mu }b$ and $\overline{c}\gamma
_{\nu }C\overline{c}^{T}$ with color triplet $\overline{\mathbf{3}_{c}}%
\otimes \mathbf{3}_{c}$ structure, where $C$ is the charge conjugation
matrix. The mass $m$ and current coupling $\Lambda $ of this tetraquark are
evaluated in the context of the QCD sum rule (SR) method \cite%
{Shifman:1978bx,Shifman:1978by}. To find width of $T$ , we employ the
three-point SR approach which is required to calculate strong couplings of $%
T $ and final state-mesons $B_{c}^{-}B_{c}^{-}$ and $B_{c}^{-}B_{c}^{\ast -}$%
: Thresholds for production of these meson pairs are below the mass of the
tetraquark $T$.

We present our results in the six sections: In Sec.\ \ref{sec:Tensor}, we
compute the mass and current coupling of $T$ using the two-point sum rule
approach. In this section, we compare our predictions with results available
in the literature. Because, the tetraquark $T$ can dissociate to $%
B_{c}^{-}B_{c}^{-}$, $B_{c}^{-}B_{c}^{\ast -}$ and $B_{c}^{\ast
-}B_{c}^{\ast -}$ mesons, we explore the decay $T\rightarrow
B_{c}^{-}B_{c}^{-}$ in Sec.\ \ref{sec:PSPSWidth}. The second channel $%
T\rightarrow B_{c}^{-}B_{c}^{\ast -}$ is analyzed in Sec.\ \ref{sec:PSVWidth}%
. The section \ref{sec:VVWidth} is devoted to investigation of the process $%
T\rightarrow B_{c}^{\ast -}B_{c}^{\ast -}$. Here, using predictions for the
partial widths of aforementioned decays, we also estimate the full width of
the tensor tetraquark $T$. The last section is reserved for our final notes.


\section{Mass and current coupling of the tensor tetraquark $bb\overline{c}%
\overline{c}$}

\label{sec:Tensor}

As it has been noted above, we model the tensor tetraquark $T=bb\overline{c}%
\overline{c}$ in the diquark-antidiquark picture as a state made of
axial-vector diquarks. The relevant interpolating current necessary for our
analysis\ in the SR framework is determined by the formula%
\begin{equation}
J_{\mu \nu }(x)=[b_{a}^{T}(x)C\gamma _{\mu }b_{b}(x)][\overline{c}%
_{a}(x)\gamma _{\nu }C\overline{c}_{b}^{T}(x)].  \label{eq:C1}
\end{equation}

The mass $m$ and current coupling $\Lambda $ of this state can be extracted
from the two-point SRs. To derive the required sum rules, we start from the
correlation function
\begin{equation}
\Pi _{\mu \nu \alpha \beta }(p)=i\int d^{4}xe^{ipx}\langle 0|\mathcal{T}%
\{J_{\mu \nu }(x)J_{\alpha \beta }^{\dag }(0)\}|0\rangle ,  \label{eq:CF1}
\end{equation}%
where the symbol $\mathcal{T}$ \ indicates the time-ordering of two $J_{\mu
\nu }(x)$ and $J_{\alpha \beta }^{\dag }(0)$ currents' product.

In accordance with paradigm of the sum rule method, the correlation function
$\Pi _{\mu \nu \alpha \beta }(p)$ has to be expressed using both the
physical parameters of the tetraquark $T$ and quark-gluon degrees of
freedom. The correlator $\Pi _{\mu \nu \alpha \beta }^{\mathrm{Phys}}(p)$ in
terms of the tetraquark's mass and current coupling has the following form
\begin{equation}
\Pi _{\mu \nu \alpha \beta }^{\mathrm{Phys}}(p)=\frac{\langle 0|J_{\mu \nu
}|T(p,\varepsilon )\rangle \langle T(p,\varepsilon )|J_{\alpha \beta }^{\dag
}|0\rangle }{m^{2}-p^{2}}+\cdots ,  \label{eq:Phys1}
\end{equation}%
which is derived by inserting into Eq.\ (\ref{eq:CF1}) a full set of states
with the quark content and quantum numbers of the tetraquark $T$ , and
integrating obtained expression over $x$. The term presented in Eq.\ (\ref%
{eq:Phys1}) is a contribution to $\Pi _{\mu \nu \alpha \beta }^{\mathrm{Phys}%
}(p)$ arising from the ground-level state: Contributions of higher
resonances and continuum states are shown by the dots.

It is useful to rewrite $\Pi _{\mu \nu \alpha \beta }^{\mathrm{Phys}}(p)$ by
employing the mass and current coupling of the tetraquark $T$. To this end,
we employ the following matrix element
\begin{equation}
\langle 0|J_{\mu \nu }|T(p,\varepsilon )\rangle =\Lambda \varepsilon _{\mu
\nu }^{(\lambda )}(p),  \label{eq:ME1}
\end{equation}%
where $\Lambda $ is the coupling of the current $J_{\mu \nu }$ to the state $%
T$, and $\varepsilon _{\mu \nu }^{(\lambda )}(p)$ is the polarization tensor
of the tetraquark $T$.

Then, the expression for $\Pi _{\mu \nu \alpha \beta }^{\mathrm{Phys}}(p)$
can be recast into a simple form%
\begin{eqnarray}
\Pi _{\mu \nu \alpha \beta }^{\mathrm{Phys}}(p) &=&\frac{\Lambda ^{2}}{%
m^{2}-p^{2}}\left\{ \frac{\widetilde{g}_{\mu \alpha }\widetilde{g}_{\nu
\beta }+\widetilde{g}_{\mu \beta }\widetilde{g}_{\nu \alpha }}{2}-\frac{%
\widetilde{g}_{\mu \nu }\widetilde{g}_{\alpha \beta }}{3}\right\} +..,
\notag \\
&&  \label{eq:Phys2}
\end{eqnarray}%
where the ellipses stand for contributions of spin-$0$ and -$1$ particles.
To obtain Eq.\ (\ref{eq:Phys2}) we have used the formula%
\begin{eqnarray}
\sum\limits_{\lambda }\varepsilon _{\mu \nu }^{(\lambda )}(p)\varepsilon
_{\alpha \beta }^{\ast (\lambda )}(p) &=&\frac{1}{2}(\widetilde{g}_{\mu
\alpha }\widetilde{g}_{\nu \beta }+\widetilde{g}_{\mu \beta }\widetilde{g}%
_{\nu \alpha })  \notag \\
&&-\frac{1}{3}\widetilde{g}_{\mu \nu }\widetilde{g}_{\alpha \beta },
\end{eqnarray}%
with%
\begin{equation}
\widetilde{g}_{\mu \nu }=-g_{\mu \nu }+\frac{p_{\mu }p_{\nu }}{p^{2}}.
\end{equation}%
The correlator $\Pi _{\mu \nu \alpha \beta }^{\mathrm{Phys}}(p)$ contains
numerous Lorentz structures. In Eq.\ (\ref{eq:Phys2}) the component $(g_{\nu
\beta }g_{\mu \alpha }+g_{\mu \beta }g_{\nu \alpha })$ receives a
contribution from only the spin-2 particle, whereas terms proportional to \ $%
g_{\mu \alpha }p_{\nu }p_{\beta }/p^{2}$ and similar ones also contain
contributions of spin-$0$ and -$1$ states. Therefore,  in our analysis we
use the structure $(g_{\mu \alpha }g_{\nu \beta }+g_{\mu \beta }g_{\nu
\alpha })$ and corresponding invariant amplitude $\Pi ^{\mathrm{Phys}}(p^{2})
$.

To get QCD side of the sum rules, it is necessary to find the correlation
function $\Pi _{\mu \nu \alpha \beta }(p)$ using the quark propagators and
compute it with some accuracy in the operator product expansion ($\mathrm{OPE%
}$). In the present paper, we calculate the correlation function by taking
into account a nonperturbative term proportional to the gluon condensate $%
\langle \alpha _{s}G^{2}/\pi \rangle $. For these purposes, we substitute
the current $J_{\mu \nu }(x)$ into Eq.\ (\ref{eq:CF1}), contract
corresponding quark fields and express obtained formula using the quark
propagators. As result, we find
\begin{eqnarray}
&&\Pi _{\mu \nu \alpha \beta }^{\mathrm{OPE}}(p)=i\int d^{4}xe^{ipx}\left\{
\mathrm{Tr}\left[ \gamma _{\nu }\widetilde{S}_{c}^{b^{\prime }b}(-x)\gamma
_{\alpha }S_{c}^{a^{\prime }a}(-x)\right] \right.   \notag \\
&&\times \left[ \mathrm{Tr}\left[ \gamma _{\beta }\widetilde{S}%
_{b}^{aa^{\prime }}(x)\gamma _{\mu }S_{b}^{bb^{\prime }}(x)\right] \right.
\notag \\
&&\left. -\mathrm{Tr}\left[ \gamma _{\beta }\widetilde{S}_{b}^{ba^{\prime
}}(x)\gamma _{\mu }S_{b}^{ab^{\prime }}(x)\right] \right] +\mathrm{Tr}\left[
\gamma _{\nu }\widetilde{S}_{c}^{a^{\prime }b}(-x)\right.   \notag \\
&&\left. \times \gamma _{\alpha }S_{c}^{b^{\prime }a}(-x)\right] \left[
\mathrm{Tr}\left[ \gamma _{\beta }\widetilde{S}_{b}^{ba^{\prime }}(x)\gamma
_{\mu }S_{b}^{ab^{\prime }}(x)\right] \right.   \notag \\
&&\left. \left. -\mathrm{Tr}\left[ \gamma _{\beta }\widetilde{S}%
_{b}^{aa^{\prime }}(x)\gamma _{\mu }S_{b}^{bb^{\prime }}(x)\right] \right]
\right\} .  \label{eq:QCD1}
\end{eqnarray}%
Here, $S_{b(c)}(x)$ are $b$ and $c$ quark propagators which can be found in
Ref.\ \cite{Agaev:2020zad}. In Eq.\ (\ref{eq:QCD1}), we have also utilized
the shorthand notation
\begin{equation}
\widetilde{S}_{b(c)}(x)=CS_{b(c)}(x)C.
\end{equation}%
In SR analysis, we are going to employ the amplitude $\Pi ^{\mathrm{OPE}%
}(p^{2})$ which corresponds in $\Pi _{\mu \nu \alpha \beta }^{\mathrm{OPE}%
}(p)$ to the structure $(g_{\mu \alpha }g_{\nu \beta }+g_{\mu \beta }g_{\nu
\alpha })$ . By this way, we avoid contaminations from the spin-$0$ and $1$
particles, because this structure appears only in spin-$2$ term and can be
extracted directly from $\Pi _{\mu \nu \alpha \beta }^{\mathrm{OPE}}(p)$.

The sum rules for $m$ and $\Lambda $ are derived by equating two amplitudes $%
\Pi ^{\mathrm{Phys}}(p^{2})$ and $\Pi ^{\mathrm{OPE}}(p^{2})$ and carrying
out prescriptions of the sum rule analysis. In other words, we apply the
Borel transformation to both sides of obtained equality and perform the
continuum subtraction. Then SRs for the mass and current coupling read%
\begin{equation}
m^{2}=\frac{\Pi ^{\prime }(M^{2},s_{0})}{\Pi (M^{2},s_{0})},  \label{eq:Mass}
\end{equation}%
and
\begin{equation}
\Lambda ^{2}=e^{m^{2}/M^{2}}\Pi (M^{2},s_{0}),  \label{eq:Coupl}
\end{equation}%
where $\Pi (M^{2},s_{0})$ is the amplitude $\Pi ^{\mathrm{OPE}}(p^{2})$
after the Borel transformation and continuum subtraction. Here, $M^{2}$ and $%
s_{0}$ are the Borel and continuum subtraction parameters, respectively:
They appear in SRs due to performed manipulations. In Eq.\ (\ref{eq:Mass})
the function $\Pi ^{\prime }(M^{2},s_{0})$ is equal to $d\Pi
(M^{2},s_{0})/d(-1/M^{2})$.

In numerical analysis, for the masses of the quarks, we use $m_{b}(\mu
=m_{b})=4.18_{-0.02}^{+0.03}~\mathrm{GeV}$ and$\ m_{c}(\mu =m_{c})=(1.27\pm
0.02)~\mathrm{GeV}$, which correspond to the running masses in the $%
\overline{\mathrm{MS}}$ scheme at the scales $\mu =m_{c}$ and $\mu =m_{b}$
\cite{PDG:2022}, respectively. For the gluon vacuum condensate, we employ $%
\langle \alpha _{s}G^{2}/\pi \rangle =(0.012\pm 0.004)~\mathrm{GeV}^{4}$
\cite{Shifman:1978bx,Shifman:1978by}.

\begin{figure}[h]
\includegraphics[width=8.5cm]{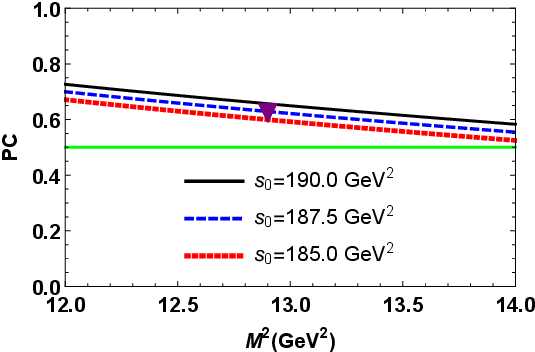}
\caption{The pole contribution $\mathrm{PC}$ vs the Borel parameter $M^{2}$
at fixed $s_{0}$. The mass $m$ of the tetraquark $T$ has been evaluated at a
position fixed by the red triangle. }
\label{fig:PC}
\end{figure}

Auxiliary quantities $M^{2}$ and $s_{0}$ are chosen inside of the regions
\begin{equation}
M^{2}\in \lbrack 12,14]~\mathrm{GeV}^{2},\ s_{0}\in \lbrack 185,190]~\mathrm{%
GeV}^{2}.  \label{eq:Wind1}
\end{equation}%
These working windows satisfy constraints of the sum rule computations.
Thus, at $M^{2}=14~\mathrm{GeV}^{2}$ and $M^{2}=12~\mathrm{GeV}^{2}$ on the
average in $s_{0}$ the pole contribution ($\mathrm{PC}$)
\begin{equation}
\mathrm{PC}=\frac{\Pi (M^{2},s_{0})}{\Pi (M^{2},\infty )},  \label{eq:PC}
\end{equation}%
is $\mathrm{PC}\approx 0.55$ and $\mathrm{PC}$ $\approx 0.7$, respectively
(Fig.\ \ref{fig:PC}). At $M^{2}=12~\mathrm{GeV}^{2}$ the nonperturbative
term is positive and forms less than $1\%$ of $\Pi (M^{2},s_{0})$.

The quantities $m$ and $\Lambda $ are estimated as mean values of these
parameters over the regions Eq.\ (\ref{eq:Wind1})
\begin{eqnarray}
m &=&(12.795\pm 0.090\pm 0.025\pm 0.018)~\mathrm{GeV},  \notag \\
\Lambda  &=&2.51_{-0.28-0.10-0.04}^{+0.28+0.06+0.03}~\mathrm{GeV}^{5}.
\label{eq:Result1}
\end{eqnarray}%
The predictions for $m$ and $\Lambda $ are effectively equal to SR results
at $M^{2}=12.9~\mathrm{GeV}^{2}$ and $s_{0}=187.5~\mathrm{GeV}^{2}$. At this
point the pole contribution is equal to $0.63$, which ensures the dominance
of $\mathrm{PC}$ and demonstrates ground-level nature of $T$. The fist and
most important errors in Eq.\ (\ref{eq:Result1}) are generated by the choice
of the parameters $M^{2}$ and $s_{0}$, the second and third ambiguities are
connected with variations of the quark masses $m_{b}$ and $m_{c}$,
respectively. The corrections $\leq |0.001|~\mathrm{GeV}$ due to
uncertainties in the gluon condensate are negligible and can be safely
neglected. Another sources of potential errors are ones due to scale
dependence of the parameters used in numerical computations. But the gluon
condensate $\langle \alpha _{s}G^{2}/\pi \rangle $ is the $\mu $-scale
independent parameter, whereas rescaling  in $m_{b}(\mu )$ and $m_{c}(\mu )$
can be treated as their uncertainties which have, as is seen from Eq.\ (\ref%
{eq:Result1}), relatively small effects on extracted quantities. Therefore,
for $m_{b}$ and $m_{c}$ throughout this work we use the $\overline{\mathrm{MS%
}}$ scheme results at fixed scales. The mass $m$ of the tetraquark $T$ as
function of the parameters $M^{2}$ and $s_{0}$ is depicted in Fig.\ \ref%
{fig:Mass}.

\begin{widetext}

\begin{figure}[h!]
\begin{center}
\includegraphics[totalheight=6cm,width=8cm]{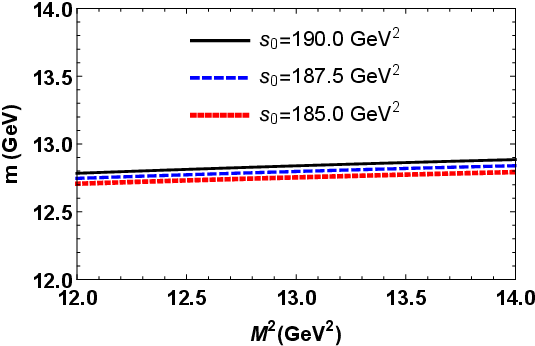}
\includegraphics[totalheight=6cm,width=8cm]{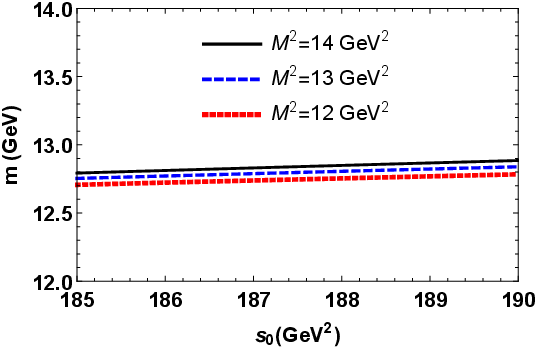}
\end{center}
\caption{Dependence of the mass $m$ on the Borel  $M^{2}$ (left panel),
and continuum threshold $s_0$ parameters (right panel).}
\label{fig:Mass}
\end{figure}

\end{widetext}

The mass of the four-quark tensor structures $bb\overline{c}\overline{c}$/$cc%
\overline{b}\overline{b}$ was calculated in the context of various methods.
Thus, using a nonrelativistic quark model the authors of Ref.\ \cite%
{Wang:2019rdo} found that the minimal \ mass of the state $bb\overline{c}%
\overline{c}$ is around of $12.868~\mathrm{GeV}$. It is interesting that in
the relativistic quark model the mass of such lowest-level tensor tetraquark
was estimated $12.859~\mathrm{GeV}$ \cite{Galkin:2023wox}. Within
uncertainties of computations these results are consistent with our
prediction.

The tetraquarks $bb\overline{c}\overline{c}$ were
considered in Ref.\ \cite{Chen:2019vrj} in a nonrelativistic chiral quark
model. Calculations carried out there suggest that no bound state can be
formed for $bb\overline{c}\overline{c}$ and $bc\overline{b}\overline{c}$
systems. The mass spectra of different all-heavy tetraquarks including $bb%
\overline{c}\overline{c}$/$cc\overline{b}\overline{b}$ ones were explored in
Ref.\ \cite{Liu:2019zuc}, as well. In this article, analysis was performed
in the framework of the potential model which includes the linear confining
potential, Coulomb potential, and spin-spin interactions. It was
demonstrated that the tensor tetraquark $bb\overline{c}\overline{c}$ has the
mass equal to $12.972~\mathrm{GeV}$. This estimate is slightly higher than
our result but is, qualitatively, in accord with all samples considered till
now. In other words, in these papers the exotic tensor meson $bb\overline{c}%
\overline{c}$ was found unstable against strong decays. Only in some
publications, in Ref.\ \cite{Wang:2021taf} for instance, this tensor
tetraquark has the mass $12.37_{-0.16}^{+0.19}~\mathrm{GeV}$ and is
strong-interaction stable structure.

Our result for $m$ implies that it is above the $B_{c}B_{c}$, $%
B_{c}B_{c}^{\ast }$, and $B_{c}^{\ast }B_{c}^{\ast }$ thresholds $12.55~%
\mathrm{GeV}$, $12.62~\mathrm{GeV}$ and $12.68~\mathrm{GeV}$ and can decay
to these final states. In the next sections, we are going to calculate
partial widths of these processes, and estimate full width of the tensor
tetraquark $bb\overline{c}\overline{c}$.


\section{Decay $T\rightarrow B_{c}^{-}B_{c}^{-}$}

\label{sec:PSPSWidth}


We start from analysis of the process $T\rightarrow B_{c}^{-}B_{c}^{-}$. The
width of this decay can be computed using the strong coupling $g$ of
particles at the vertex $TB_{c}^{-}B_{c}^{-}$. In order to find $g$ it is
convenient to analyze the three-point correlation function
\begin{eqnarray}
\Pi _{\mu \nu }(p,p^{\prime }) &=&i^{2}\int d^{4}xd^{4}ye^{ip^{\prime
}y}e^{-ipx}\langle 0|\mathcal{T}\{J^{B_{c}}(y)  \notag \\
&&\times J^{B_{c}}(0)J_{\mu \nu }^{\dagger }(x)\}|0\rangle ,  \label{eq:CF3}
\end{eqnarray}%
where
\begin{equation}
J^{B_{c}}(x)=\overline{c}_{i}(x)i\gamma _{5}b_{i}(x),  \label{eq:CR3}
\end{equation}%
is the interpolating current of the pseudoscalar meson $B_{c}^{-}$.

Investigation of the correlation function $\Pi _{\mu \nu }(p,p^{\prime })$
makes it possible to derive the SR for the form factor $g(q^{2})$ that at
the mass shell $q^{2}=m_{B_{c}}^{2}$ is equal to the strong coupling $g$. To
find SR for the form factor $g(q^{2})$, we first write down $\Pi _{\mu \nu
}(p,p^{\prime })$ using the parameters of the tetraquark $T$ and meson $%
B_{c}^{-}$. The correlation function $\Pi _{\mu \nu }^{\mathrm{Phys}%
}(p,p^{\prime })$ calculated by this method forms the physical side of the
SR for the form factor $g_{1}(q^{2})$ and is given by the formula
\begin{eqnarray}
&&\Pi _{\mu \nu }^{\mathrm{Phys}}(p,p^{\prime })=\frac{\langle
0|J^{B_{c}}|B_{c}(p^{\prime })\rangle }{p^{\prime 2}-m_{B_{c}}^{2}}\frac{%
\langle 0|J^{B_{c}}|B_{c}(q)\rangle }{q^{2}-m_{B_{c}}^{2}}  \notag \\
&&\times \langle B_{c}(p^{\prime })B_{c}(q)|T(p,\varepsilon )\rangle \frac{%
\langle T(p,\varepsilon )|J_{\mu \nu }^{\dagger }|0\rangle }{p^{2}-m^{2}}%
+\cdots ,  \notag \\
&&  \label{eq:CF5}
\end{eqnarray}%
with $m_{B_{c}}=(6274.47\pm 0.27)~\mathrm{MeV}$ being the mass of $B_{c}^{-}$
meson \cite{PDG:2022}. The correlation function $\Pi _{\mu \nu }^{\mathrm{%
Phys}}(p,p^{\prime })$ is found after isolating an effect of the
ground-level particles, whereas contributions of higher and continuum states
are denoted by the ellipses. To recast this correlator into a form suitable
for further analysis, we use the matrix element
\begin{equation}
\langle 0|J^{B_{c}}|B_{c}\rangle =\frac{f_{B_{c}}m_{B_{c}}^{2}}{m_{b}+m_{c}},
\label{eq:ME2}
\end{equation}%
where $f_{B_{c}}=(476\pm 27)~\mathrm{MeV}$ is the decay constant of the
mesons $B_{c}^{\pm }$ \cite{Veliev:2010vd}. We model the vertex of the
tensor tetraquark and two pseudoscalar mesons by the formula%
\begin{equation}
\langle B_{c}(p^{\prime })B_{c}(q)|T(p,\varepsilon )\rangle
=-g(q^{2})\varepsilon _{\alpha \beta }^{(\lambda )}(p)p^{\prime \alpha
}q^{\beta },  \label{eq:ME3}
\end{equation}%
with $g(q^{2})$ being the strong form factor that at the mass shell $%
q^{2}=m_{B_{c}}^{2}$ determines the coupling $g$ of interest. By taking into
account $q=p-p^{\prime }$ and
\begin{equation}
\varepsilon _{\alpha \beta }^{(\lambda )}(p)p^{\alpha }=\varepsilon _{\alpha
\beta }^{(\lambda )}(p)p^{\beta }=0,
\end{equation}%
we get%
\begin{equation}
\langle B_{c}(p^{\prime })B_{c}(q)|T(p,\varepsilon )\rangle
=g(q^{2})\varepsilon _{\alpha \beta }^{(\lambda )}(p)p^{\prime \alpha
}p^{\prime \beta }.
\end{equation}%
For the correlator, we find
\begin{eqnarray}
&&\Pi _{\mu \nu }^{\mathrm{Phys}}(p,p^{\prime })=\frac{g(q^{2})\Lambda
f_{B_{c}}^{2}m_{B_{c}}^{4}}{(m_{b}+m_{c})^{2}\left( p^{2}-m^{2}\right)
(p^{\prime 2}-m_{B_{c}}^{2})}  \notag \\
&&\times \frac{1}{(q^{2}-m_{B_{c}}^{2})}\left\{ \frac{\lambda ^{2}}{3}g_{\mu
\nu }+\left[ \frac{m_{B_{c}}^{2}}{m^{2}}+\frac{2\lambda ^{2}}{3m^{2}}\right]
p_{\mu }p_{\nu }\right.  \notag \\
&&\left. +p_{\mu }^{\prime }p_{\nu }^{\prime }-\frac{%
m^{2}+m_{B_{c}}^{2}-q^{2}}{2m^{2}}(p_{\mu }p_{\nu }^{\prime }+p_{\nu }p_{\mu
}^{\prime })\right\} ,
\end{eqnarray}%
where $\lambda =\lambda (m,m_{B_{c}},q)$ and
\begin{equation}
\lambda (x,y,z)=\frac{\sqrt{%
x^{4}+y^{4}+z^{4}-2(x^{2}y^{2}+x^{2}z^{2}+y^{2}z^{2})}}{2x}.
\end{equation}%
The correlation function $\Pi _{\mu \nu }^{\mathrm{Phys}}(p,p^{\prime })$
has the structure
\begin{eqnarray}
&&\Pi _{\mu \nu }^{\mathrm{Phys}}(p,p^{\prime })=\Pi _{0}^{\mathrm{Phys}%
}g_{\mu \nu }+\Pi _{1}^{\mathrm{Phys}}p_{\mu }p_{\nu }+\Pi _{2}^{\mathrm{Phys%
}}p_{\mu }^{\prime }p_{\nu }^{\prime }  \notag \\
&&+\Pi _{3}^{\mathrm{Phys}}p_{\mu }p_{\nu }^{\prime }+\Pi _{4}^{\mathrm{Phys}%
}p_{\nu }p_{\mu }^{\prime }.  \label{eq:PhysSide}
\end{eqnarray}%
The amplitudes $\Pi _{i}^{\mathrm{Phys}}$ are functions of the variables $%
p^{2},p^{\prime 2},q^{2}$, and depend on the parameters $m^{2}$ and $%
m_{B_{c}}^{2}$ which are omitted for compactness of the expression. To
derive the SR for the form factor $g(q^{2})$, we work with the structure $%
p_{\mu }^{\prime }p_{\nu }^{\prime }$ and corresponding invariant amplitude $%
\Pi ^{\mathrm{Phys}}(p^{2},p^{\prime 2},q^{2})$ (we have omitted a subscript
$2$). It has the form%
\begin{eqnarray}
&&\Pi ^{\mathrm{Phys}}(p^{2},p^{\prime 2},q^{2})=\frac{g(q^{2})\Lambda
f_{B_{c}}^{2}m_{B_{c}}^{4}}{(m_{b}+m_{c})^{2}\left( p^{2}-m^{2}\right) }
\notag \\
&&\times \frac{1}{(p^{\prime 2}-m_{B_{c}}^{2})(q^{2}-m_{B_{c}}^{2})}.
\end{eqnarray}

The QCD side of the sum rule, i.e., the correlation function $\Pi _{\mu \nu
}^{\mathrm{OPE}}(p,p^{\prime })$ is determined by the formula%
\begin{eqnarray}
&&\Pi _{\mu \nu }^{\mathrm{OPE}}(p,p^{\prime })=2i^{2}\int
d^{4}xd^{4}ye^{ip^{\prime }y}e^{-ipx}\left\{ \mathrm{Tr}\left[ \gamma
_{5}S_{b}^{ia}(y-x)\right. \right.  \notag \\
&&\left. \times \gamma _{\nu }\widetilde{S}_{b}^{jb}(-x)\gamma _{5}%
\widetilde{S}_{c}^{aj}(x)\gamma _{\mu }S_{c}^{bi}(x-y)\right]  \notag \\
&&\left. -\mathrm{Tr}\left[ \gamma _{5}S_{b}^{ia}(y-x)\gamma _{\nu }%
\widetilde{S}_{b}^{jb}(-x)\gamma _{5}\widetilde{S}_{c}^{bj}(x)\gamma _{\mu
}S_{c}^{ai}(x-y)\right] \right\} .  \notag \\
&&  \label{eq:CF6}
\end{eqnarray}%
The correlator $\Pi _{\mu \nu }^{\mathrm{OPE}}(p,p^{\prime })$ has the same
Lorentz structure as $\Pi _{\mu \nu }^{\mathrm{Phys}}$. We label by $\Pi ^{%
\mathrm{OPE}}(p^{2},p^{\prime 2},q^{2})$ the amplitude of the structure $%
p_{\mu }^{\prime }p_{\nu }^{\prime }$ and employ it in the following
analysis. Having equated functions $\Pi ^{\mathrm{Phys}}(p^{2},p^{\prime
2},q^{2})$ and $\Pi ^{\mathrm{OPE}}(p^{2},p^{\prime 2},q^{2})$, and
fulfilled the Borel transformations over variables $-p^{2}$ and $-p^{\prime
2}$ and continuum subtraction procedures, we find
\begin{eqnarray}
&&g(q^{2})=\frac{(m_{b}+m_{c})^{2}}{\Lambda f_{B_{c}}^{2}m_{B_{c}}^{4}}%
(q^{2}-m_{B_{c}}^{2})  \notag \\
&&\times e^{m^{2}/M_{1}^{2}}e^{m_{B_{c}}^{2}/M_{2}^{2}}\Pi (\mathbf{M}^{2},%
\mathbf{s}_{0},q^{2}),  \label{eq:SRCoup2}
\end{eqnarray}%
where
\begin{eqnarray}
&&\Pi (\mathbf{M}^{2},\mathbf{s}_{0},q^{2})=\int_{4\mathcal{M}%
^{2}}^{s_{0}}ds\int_{\mathcal{M}^{2}}^{s_{0}^{\prime }}ds^{\prime }\rho
(s,s^{\prime },q^{2})  \notag \\
&&\times e^{-s/M_{1}^{2}}e^{-s^{\prime }/M_{2}^{2}},  \label{eq:InvAmp}
\end{eqnarray}%
and $\mathcal{M}=m_{b}+m_{c}$. Here, $\Pi (\mathbf{M}^{2},\mathbf{s}%
_{0},q^{2})$ is the Borel transformed and continuum subtracted amplitude $%
\Pi ^{\mathrm{OPE}}(p^{2},p^{\prime 2},q^{2})$ corresponding to the term $%
\sim p_{\mu }^{\prime }p_{\nu }^{\prime }$. The spectral density $\rho
(s,s^{\prime },q^{2})$ is calculated as a imaginary part of the component $%
\sim p_{\mu }^{\prime }p_{\nu }^{\prime }$ in the correlation function. The $%
(\mathbf{M}^{2},\mathbf{s}_{0})$ in Eq.\ (\ref{eq:InvAmp}) are two pairs of
the parameters: The parameters $(M_{1}^{2},s_{0})$ corresponds to the tensor
tetraquark channel, and in numerical analysis we use Eq.\ (\ref{eq:Wind1}).
The pair $(M_{2}^{2},s_{0}^{\prime })$ describes the $B_{c}^{-}$ channel for
which we employ
\begin{equation}
M_{2}^{2}\in \lbrack 6.5,7.5]~\mathrm{GeV}^{2},\ s_{0}^{\prime }\in \lbrack
45,47]~\mathrm{GeV}^{2}.  \label{eq:Wind3}
\end{equation}

The strong coupling $g$ is equal to the form factor $g(m_{B_{c}}^{2})$
computed at the mass shell $q^{2}=m_{B_{c}}^{2}$. Because the SR method is
applicable only in the domain $q^{2}<0$, to get $g(m_{B_{c}}^{2})$ one has
to extrapolate the QCD numerical data to a region of positive $q^{2}$. For
these purposes, it is appropriate to use a variable $Q^{2}=-q^{2}$, and
denote the fit function $\mathcal{F}(Q^{2})$, which at $Q^{2}>0$ gives
results equal to SR data, but can be easily extrapolated to $Q^{2}<0$. One
of possible choices for the such function is
\begin{equation}
\mathcal{F}(Q^{2})=\mathcal{F}^{0}\mathrm{\exp }\left[ c_{1}\frac{Q^{2}}{%
m^{2}}+c_{2}\left( \frac{Q^{2}}{m^{2}}\right) ^{2}\right] .  \label{eq:FitF}
\end{equation}%
This function contains the unknown parameters $\mathcal{F}^{0}$, $c_{1}$,
and $c_{2}$ which should be determined from a fitting procedure.

\begin{figure}[h]
\includegraphics[width=8.5cm]{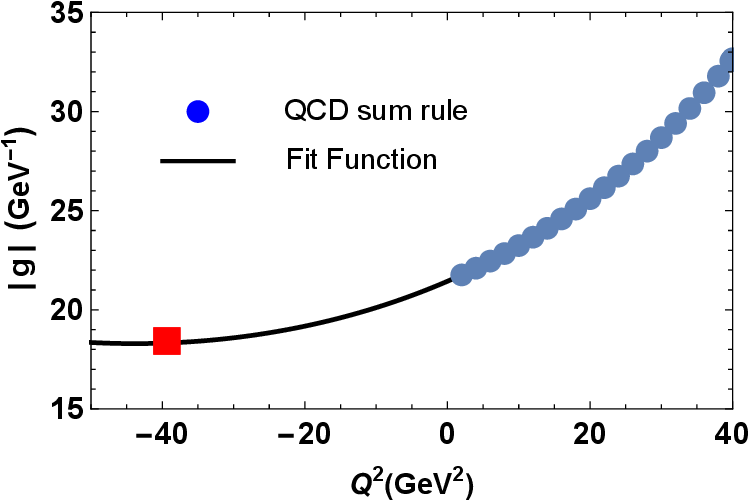}
\caption{The QCD data for the form factor $|g(Q^{2})|$ and the extrapolating
function $\mathcal{F}$. The red square shows the point $Q^{2}=-m_{B_{c}}^{2}$%
. }
\label{fig:Fit}
\end{figure}

We carry out the numerical computations varying $Q^{2}$ inside of the
interval $Q^{2}=1-40~\mathrm{GeV}^{2}$. The QCD data extracted from the SR
analysis for the form factor $|g(Q^{2})|$ are plotted in Fig.\ \ref{fig:Fit}%
. Having compared these data and Eq.\ (\ref{eq:FitF}), one can obtain the
parameters $\mathcal{F}^{0}=21.44~\mathrm{GeV}^{-1}$, $c_{1}=1.18$, and $%
c_{2}=2.21$ of the function $\mathcal{F}(Q^{2})$. This function is plotted
in Fig.\ \ref{fig:Fit} as well: A reasonable agreement between $\mathcal{F}%
(Q^{2})$ and QCD data is evident.

For the strong coupling $g$, we find
\begin{equation}
|g|\equiv \mathcal{F}%
(-m_{B_{c}}^{2})=18.33_{-2.02-0.59-0.35}^{+2.02+0.59+0.79}\ \mathrm{GeV}%
^{-1}.  \label{eq:Coupl1}
\end{equation}%
Here, the errors in $|g|$ are generated by the choice of $M^{2}$ and $s_{0}$%
, and ambiguities in $m_{b}$ and $m_{c}$, respectively. The partial width of
the decay $T\rightarrow B_{c}^{-}B_{c}^{-}$ is determined by the following
formula%
\begin{equation}
\Gamma \left[ T\rightarrow B_{c}^{-}B_{c}^{-}\right] =\frac{g^{2}\lambda }{%
2\cdot 960\pi m^{2}}\left( m^{2}-2m_{B_{c}}^{2}\right) ^{2},  \label{eq:PW1}
\end{equation}%
with $\lambda $ being equal to $\lambda (m,m_{B_{c}},m_{B_{c}})$. Having
applied Eqs.\ (\ref{eq:Coupl1}) and (\ref{eq:PW1}) and other input
parameters, we get the partial width of this decay
\begin{equation}
\Gamma \left[ T\rightarrow B_{c}^{-}B_{c}^{-}\right] =16.3_{-4.06}^{+4.35}~%
\mathrm{MeV},  \label{eq:DW1}
\end{equation}%
where we present the total errors of calculations.

\section{Decay $T\rightarrow B_{c}^{-}B_{c}^{\ast -}$}

\label{sec:PSVWidth}


Here, we study the process $T\rightarrow B_{c}^{-}B_{c}^{\ast -}$ and
calculate its partial width. To this end, we begin from analysis of the
correlation function
\begin{eqnarray}
\Pi _{\alpha \mu \nu }(p,p^{\prime }) &=&i^{2}\int d^{4}xd^{4}ye^{ip^{\prime
}y}e^{-ipx}\langle 0|\mathcal{T}\{J_{\alpha }^{B_{c}^{\ast }}(y)  \notag \\
&&\times J^{B_{c}}(0)J_{\mu \nu }^{\dagger }(x)\}|0\rangle ,  \label{eq:CF6a}
\end{eqnarray}%
which is necessary to evaluate the strong coupling at the vertex $%
TB_{c}B_{c}^{\ast }$. In Eq.\ (\ref{eq:CF6a}) $J_{\mu }^{B_{c}^{\ast }}(y)$
is the interpolating current of the vector meson $B_{c}^{\ast -}$
\begin{equation}
J_{\mu }^{B_{c}^{\ast }}(y)=\overline{c}_{i}(x)\gamma _{\mu }b_{i}(x).
\end{equation}

The SR investigation requires calculation the correlator $\Pi _{\alpha \mu
\nu }(p,p^{\prime })$ in two regions. First, one should find it using the
physical parameters of particles involved into the decay process. This
expression which contains contribution of only ground-state particles is
\begin{eqnarray}
&&\Pi _{\alpha \mu \nu }^{\mathrm{Phys}}(p,p^{\prime })=\frac{\langle
0|J_{\alpha }^{B_{c}^{\ast }}|B_{c}^{\ast }(p^{\prime },\epsilon )\rangle }{%
p^{\prime 2}-m_{B_{c}^{\ast }}^{2}}\frac{\langle 0|J^{B_{c}}|B_{c}(q)\rangle
}{q^{2}-m_{B_{c}}^{2}}  \notag \\
&&\times \langle B_{c}^{\ast }(p^{\prime },\epsilon
)B_{c}(q)|T(p,\varepsilon )\rangle \frac{\langle T(p,\varepsilon )|J_{\mu
\nu }^{\dagger }|0\rangle }{p^{2}-m^{2}}+\cdots .  \notag \\
&&  \label{eq:PhysSide1}
\end{eqnarray}%
The matrix element of the vector particle is
\begin{equation}
\langle 0|J_{\mu }^{B_{c}^{\ast }}|B_{c}^{\ast }(p)\rangle =f_{B_{c}^{\ast
}}m_{B_{c}^{\ast }}\epsilon _{\mu }(p),  \label{eq:ME4}
\end{equation}%
where $m_{B_{c}^{\ast }}$ and $f_{B_{c}^{\ast }}$ are the mass and decay
constant of the meson $B_{c}^{\ast }(p)$, respectively. The vertex $%
TB_{c}B_{c}^{\ast }$ is modeled by the formula
\begin{equation}
\langle B_{c}^{\ast }(p^{\prime },\epsilon )B_{c}(q)|T(p,\varepsilon
)\rangle =h(q^{2})\varepsilon _{\beta \tau \rho \delta }p^{\beta
}\varepsilon _{(\lambda )}^{\tau \sigma }(p)q_{\sigma }p^{\prime \rho
}{}\epsilon ^{\ast \delta }.  \label{eq:TPVvertex}
\end{equation}%
Having substituted these matrix elements in Eq.\ (\ref{eq:PhysSide1}), we
get
\begin{eqnarray}
&&\Pi _{\alpha \mu \nu }^{\mathrm{Phys}}(p,p^{\prime })=\frac{%
h(q^{2})\Lambda f_{B_{c}^{\ast }}m_{B_{c}^{\ast }}f_{B_{c}}m_{B_{c}}^{2}}{%
2(m_{b}+m_{c})\left( p^{2}-m^{2}\right) (p^{\prime 2}-m_{B_{c}^{\ast }}^{2})}
\notag \\
&&\times \frac{1}{(q^{2}-m_{B_{c}}^{2})}\left\{ \varepsilon
_{\lambda \beta \alpha \mu }p^{\lambda }p^{\prime \beta }p_{\nu }^{\prime
}+\varepsilon _{\lambda \beta \alpha \nu }p^{\lambda }p^{\prime \beta
}p_{\mu }^{\prime }\right.  \notag \\
&&\left. -\frac{m^{2}+m_{B_{c}^{\ast }}^{2}-q^{2}}{2m^{2}}\left( \varepsilon
_{\lambda \beta \alpha \nu }p^{\lambda }p^{\prime \beta }p_{\mu
}+\varepsilon _{\lambda \beta \alpha \mu }p^{\lambda }p^{\prime \beta
}p_{\nu }\right) \right\} .  \notag \\
&&  \label{eq:PhysSide1a}
\end{eqnarray}%
The QCD side of the sum rule is determined by the expression%
\begin{eqnarray}
&&\Pi _{\alpha \mu \nu }^{\mathrm{OPE}}(p,p^{\prime })=2i\int
d^{4}xd^{4}ye^{ip^{\prime }y}e^{-ipx}\left\{ \mathrm{Tr}\left[ \gamma
_{\alpha }S_{b}^{ia}(y-x)\right. \right.  \notag \\
&&\left. \times \gamma _{\nu }\widetilde{S}_{b}^{jb}(-x)\gamma _{5}%
\widetilde{S}_{c}^{aj}(x)\gamma _{\mu }S_{c}^{bi}(x-y)\right]  \notag \\
&&\left. -\mathrm{Tr}\left[ \gamma _{\alpha }S_{b}^{ia}(y-x)\gamma _{\nu }%
\widetilde{S}_{b}^{jb}(-x)\gamma _{5}\widetilde{S}_{c}^{bj}(x)\gamma _{\mu
}S_{c}^{ai}(x-y)\right] \right\} .  \notag \\
&&  \label{eq:QCDSide}
\end{eqnarray}%
We derive SR for the form factor $h(q^{2})$ using the first structure in
Eq.\ (\ref{eq:PhysSide1a}) and its counterpart in $\Pi _{\alpha \mu \nu }^{%
\mathrm{OPE}}(p,p^{\prime })$%
\begin{eqnarray}
h(q^{2}) &=&\frac{2(m_{b}+m_{c})(q^{2}-m_{B_{c}^{\ast }}^{2})}{\Lambda
f_{B_{c}^{\ast }}m_{B_{c}^{\ast }}f_{B_{c}}m_{B_{c}}^{2}}  \notag \\
&&\times e^{m^{2}/M_{1}^{2}}e^{m_{B_{c}^{\ast }}^{2}/M_{2}^{2}}\widehat{\Pi }%
(\mathbf{M}^{2},\mathbf{s}_{0},q^{2}),  \label{eq:SR}
\end{eqnarray}%
where $\widehat{\Pi }(\mathbf{M}^{2},\mathbf{s}_{0},q^{2})$ is the amplitude
corresponding to the structure $\varepsilon _{\lambda \beta \alpha \mu
}p^{\lambda }p^{\prime \beta }p_{\nu }^{\prime }$ in the correlation
function $\Pi _{\alpha \mu \nu }^{\mathrm{OPE}}(p,p^{\prime })$ after the
Borel transformations and continuum subtractions.

\begin{figure}[h]
\includegraphics[width=8.5cm]{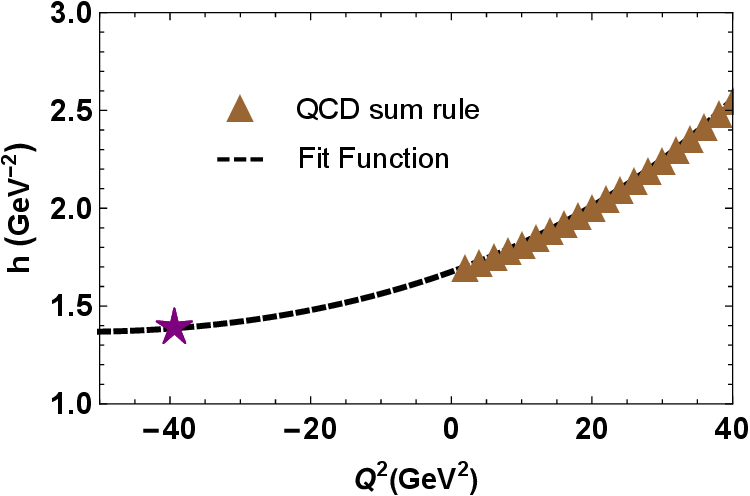}
\caption{The SR output and interpolating function for the form factor $%
h(Q^{2})$. The star shows the point $Q^{2}=-m_{B_{c}}^{2}$. }
\label{fig:FitA}
\end{figure}

In numerical computations, we utilize the theoretical predictions for the
mass and decay constant of the vector meson $B_{c}^{\ast }$
\begin{equation}
m_{B_{c}^{\ast }}=6338~\mathrm{MeV,\ }\,f_{B_{c}^{\ast }}=471~\mathrm{MeV}
\end{equation}%
from Refs.\ \cite{Godfrey:2004ya,Eichten:2019gig}, respectively. The
parameters in the $B_{c}^{\ast }$ channel are chosen in accordance with Eq.\
(\ref{eq:Wind2}):%
\begin{equation}
M_{2}^{2}\in \lbrack 6.5,7.5]~\mathrm{GeV}^{2},\ s_{0}^{\prime }\in \lbrack
49,51]~\mathrm{GeV}^{2}.  \label{eq:Wind2}
\end{equation}%
The results of numerical analysis for $h(q^{2})$ and extrapolating function $%
\widehat{\mathcal{F}}(Q^{2})$ with parameters $\widehat{\mathcal{F}}%
^{0}=1.67~\mathrm{GeV}^{-2}$, $c_{1}=1.26$, and $c_{2}=1.95$ are drawn in
Fig.\ \ref{fig:FitA}.

Having employed the fit function $\widehat{\mathcal{F}}(Q^{2})$, it is not
difficult to estimate the strong coupling $h$%
\begin{equation}
h\equiv \widehat{\mathcal{F}}(-m_{B_{c}}^{2})=(1.39\pm 0.16)\
\mathrm{GeV}^{-2}.  \label{eq:Coupl2}
\end{equation}%
The partial width of the decay $T\rightarrow B_{c}^{-}B_{c}^{\ast -}$ is
equal to
\begin{eqnarray}
\Gamma \left[ T\rightarrow B_{c}^{-}B_{c}^{\ast -}\right] &=&\frac{h^{2}%
\widehat{\lambda }}{640\pi m^{4}}\left[ m_{B_{c}}^{4}+(m^{2}-m_{B_{c}^{\ast
}}^{2})^{2}\right. ,  \notag \\
&&\left. -2m_{B_{c}}^{2}(m^{2}+m_{B_{c}^{\ast }}^{2})\right] ^{2},
\label{eq:DW1a}
\end{eqnarray}%
where $\widehat{\lambda }=\lambda (m,m_{B_{c}^{\ast }},m_{B_{c}})$. Having
applied Eqs.\ (\ref{eq:Coupl2}) and (\ref{eq:DW1a}) and other input
parameters, we evaluate the partial width of this process
\begin{equation}
\Gamma \left[ T\rightarrow B_{c}^{-}B_{c}^{\ast -}\right] =(25.2\pm 7.3)~%
\mathrm{MeV}.  \label{eq:DW1b}
\end{equation}


\section{Decay $T\rightarrow B_{c}^{\ast -}B_{c}^{\ast -}$}

\label{sec:VVWidth}


In the case of the decay $T\rightarrow B_{c}^{\ast -}B_{c}^{\ast -}$ the SR
for the strong form factor $G(q^{2})$ at the vertex $TB_{c}^{\ast
}B_{c}^{\ast }$ can be obtained from the correlation function
\begin{eqnarray}
\Pi _{\alpha \beta \mu \nu }(p,p^{\prime }) &=&i^{2}\int
d^{4}xd^{4}ye^{ip^{\prime }y}e^{-ipx}\langle 0|\mathcal{T}\{J_{\alpha
}^{B_{c}^{\ast }}(y)  \notag \\
&&\times J_{\beta }^{B_{c}^{\ast }}(0)J_{\mu \nu }^{\dagger }(x)\}|0\rangle .
\label{eq:CF7}
\end{eqnarray}

The correlation function $\Pi _{\alpha \beta \mu \nu }(p,p^{\prime })$ can
be expressed using the physical parameters of the particles involved onto
the decay process
\begin{eqnarray}
&&\Pi _{\alpha \beta \mu \nu }^{\mathrm{Phys}}(p,p^{\prime })=\frac{\langle
0|J_{\alpha }^{B_{c}^{\ast }}|B_{c}^{\ast }(p^{\prime },\epsilon
_{1})\rangle }{p^{\prime 2}-m_{B_{c}^{\ast }}^{2}}\frac{\langle 0|J_{\beta
}^{B_{c}^{\ast }}|B_{c}^{\ast }(q,\epsilon _{2})\rangle }{%
q^{2}-m_{B_{c}^{\ast }}^{2}}  \notag \\
&&\times \langle B_{c}^{\ast }(p^{\prime },\epsilon _{1})B_{c}^{\ast
}(q,\epsilon _{2})|T(p,\varepsilon )\rangle \frac{\langle T(p,\varepsilon
)|J_{\mu \nu }^{\dagger }|0\rangle }{p^{2}-m^{2}}+\cdots ,  \notag \\
&&
\end{eqnarray}%
where $\epsilon _{1\mu }$ and $\epsilon _{2\mu }$ are the polarization
vector of the $B_{c}^{\ast }$ mesons with momenta $p^{\prime }$ and $q$,
respectively. In the correlator $\Pi _{\alpha \beta \mu \nu }^{\mathrm{Phys}%
}(p,p^{\prime })$ only contribution of the ground-level particles is
presented explicitly: Contributions arising from the higher resonances and
continuum states are shown by the dots.

The correlation function can be further simplified using the matrix element
of the vertex $\langle B_{c}^{\ast }(p^{\prime },\epsilon _{1})B_{c}^{\ast
}(q,\epsilon _{2})|T(p,\varepsilon )\rangle $. In general, a
tensor-vector-vector vertex contains three independent form factors which
describe a pair of vector meson with helicities $\lambda =0$, $\pm 1$ and $%
\pm 2$ \cite{Braun:2000cs,Aliev:2018kry}. In the case of two photon decays
of a tensor meson it was argued that a dominant contribution comes from a $%
\lambda =2$ state \cite{Singer:1983bu}. In the present work, we assume that
this conclusion may be applied to decay $T\rightarrow B_{c}^{\ast
-}B_{c}^{\ast -}$, as well, and choose the vertex in the form
\begin{eqnarray}
&&\langle B_{c}^{\ast }(p^{\prime },\epsilon _{1})B_{c}^{\ast }(q,\epsilon
_{2})|T(p,\varepsilon )\rangle =G(q^{2})\varepsilon _{\tau \rho }^{(\lambda
)}\left[ (\epsilon _{1}^{\ast }\cdot q)\epsilon _{2}^{\tau \ast }p^{\prime
\rho }\right.  \notag \\
&&\left. +(\epsilon _{2}^{\ast }\cdot p^{\prime })\epsilon _{1}^{\ast \tau
}q^{\rho }-(p^{\prime }\cdot q)\epsilon _{1}^{\tau \ast }\epsilon _{2}^{\rho
\ast }-(\epsilon _{1}^{\ast }\cdot \epsilon _{2}^{\ast })p^{\prime \tau
}q^{\rho }\right] ,  \notag \\
&&  \label{eq:TVVVertex}
\end{eqnarray}%
which corresponds to a pure $\lambda =2$ final state.

As a result, for $\Pi _{\alpha \beta \mu \nu }^{\mathrm{Phys}}(p,p^{\prime
}) $ we get the lengthy expression
\begin{eqnarray}
&&\Pi _{\alpha \beta \mu \nu }^{\mathrm{Phys}}(p,p^{\prime })=\frac{%
G(q^{2})\Lambda f_{B_{c}^{\ast }}^{2}m_{B_{c}^{\ast }}^{2}}{\left(
p^{2}-m^{2}\right) (p^{\prime 2}-m_{B_{c}^{\ast }}^{2})(q^{2}-m_{B_{c}^{\ast
}}^{2})}  \notag \\
&&\times \left\{ \frac{m_{B_{c}^{\ast }}^{4}-2m_{B_{c}^{\ast
}}^{2}(2m^{2}+q^{2})+(m^{2}-q^{2})(3m^{2}-q^{2})}{12m^{2}}\right.  \notag \\
&&\times g_{\mu \nu }g_{\alpha \beta }+\frac{1}{3}g_{\mu \nu }\left( \frac{%
m_{B_{c}^{\ast }}^{2}}{m^{2}}p_{\alpha }p_{\beta }+2p_{\alpha }^{\prime
}p_{\beta }^{\prime }\right) -\frac{1}{6m^{2}}g_{\mu \nu }  \notag \\
&&\times \left[ (m_{B_{c}^{\ast }}^{2}+3m^{2}-q^{2})p_{\alpha }p_{\beta
}^{\prime }+(m_{B_{c}^{\ast }}^{2}+m^{2}-q^{2})p_{\alpha }^{\prime }p_{\beta
}\right]  \notag \\
&&+\frac{1}{2}p_{\alpha }p_{\mu }^{\prime }g_{\beta \nu }+\frac{1}{2m^{2}}%
\left( p_{\beta }p_{\nu }p_{\alpha }^{\prime }p_{\mu }^{\prime }+p_{\beta
}p_{\mu }p_{\alpha }^{\prime }p_{\nu }^{\prime }\right)  \notag \\
&&\left. +\text{other structures}\right\} .  \label{eq:PhysSide2}
\end{eqnarray}

For the QCD side of the sum rule, we obtain
\begin{eqnarray}
&&\Pi _{\alpha \beta \mu \nu }^{\mathrm{OPE}}(p,p^{\prime })=2i^{2}\int
d^{4}xd^{4}ye^{ip^{\prime }y}e^{-ipx}\left\{ \mathrm{Tr}\left[ \gamma
_{\alpha }S_{b}^{ia}(y-x)\right. \right.  \notag \\
&&\left. \times \gamma _{\nu }\widetilde{S}_{b}^{jb}(-x)\gamma _{\beta }%
\widetilde{S}_{c}^{aj}(x)\gamma _{\mu }S_{c}^{bi}(x-y)\right]  \notag \\
&&\left. -\mathrm{Tr}\left[ \gamma _{\alpha }S_{b}^{ia}(y-x)\gamma _{\nu }%
\widetilde{S}_{b}^{jb}(-x)\gamma _{\beta }\widetilde{S}_{c}^{bj}(x)\gamma
_{\mu }S_{c}^{ai}(x-y)\right] \right\} .  \notag \\
&&  \label{eq:CF8}
\end{eqnarray}%
To get the required SR for the form factor $G(q^{2})$ we utilize the
structure proportional to $p_{\alpha }p_{\mu }^{\prime }g_{\beta \nu }$ and
corresponding amplitudes $\widetilde{\Pi }^{\mathrm{Phys}}(p^{2},p^{\prime
2},q^{2})$ and $\widetilde{\Pi }^{\mathrm{OPE}}(p^{2},p^{\prime 2},q^{2})$
in both versions of the correlation function $\Pi _{\alpha \beta \mu \nu
}(p,p^{\prime })$. Then, after standard operations the sum rule for $%
G(q^{2}) $ reads
\begin{equation}
G(q^{2})=\frac{2(q^{2}-m_{B_{c}^{\ast }}^{2})}{\Lambda f_{B_{c}^{\ast
}}^{2}m_{B_{c}^{\ast }}^{2}}e^{m^{2}/M_{1}^{2}}e^{m_{B_{c}^{\ast
}}^{2}/M_{2}^{2}}\widetilde{\Pi }(\mathbf{M}^{2},\mathbf{s}_{0},q^{2}).
\label{eq:SRG}
\end{equation}

\begin{figure}[h]
\includegraphics[width=8.5cm]{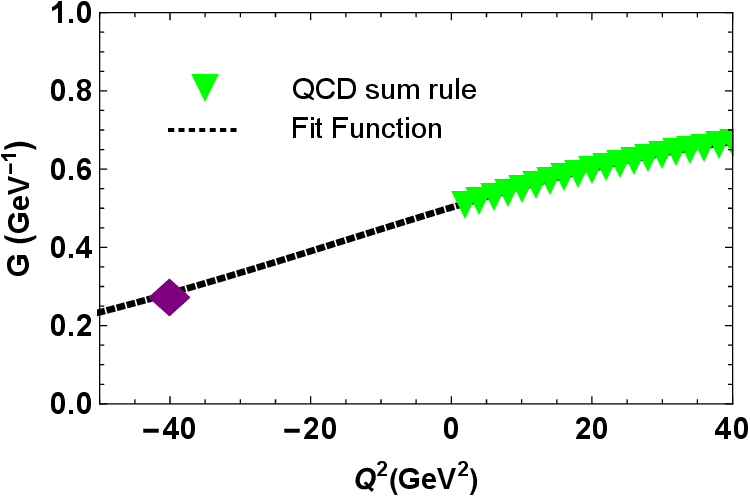}
\caption{The QCD data for the form factor $G(Q^{2})$ and fit function $%
\widetilde{\mathcal{F}}$. The diamond fixes the point $Q^{2}=-m_{B_{c}}^{%
\ast 2}$. }
\label{fig:Fit1}
\end{figure}

The results obtained for $G(Q^{2})$ are plotted in Fig.\ \ref{fig:Fit1},
where $Q^{2}$ varies inside of limits $Q^{2}=1-40~\mathrm{GeV}^{2}$. Then,
having confronted QCD output and Eq.\ (\ref{eq:FitF}), it is easy to find $%
\widetilde{\mathcal{F}}^{0}=0.50~\mathrm{GeV}^{-1}$, $c_{1}=1.77$, and $%
c_{2}=-2.39$ in the function $\widetilde{\mathcal{F}}(Q^{2})$. It is also
plotted in Fig.\ \ref{fig:Fit1}, where one sees a nice agreement of $%
\widetilde{\mathcal{F}}(Q^{2})$ and QCD data.

For the strong coupling $G$, we find
\begin{equation}
G\equiv \widetilde{\mathcal{F}}(-m_{B_{c}^{\ast
}}^{2})=0.28_{-0.04}^{+0.06}\ \mathrm{GeV}^{-1}.
\end{equation}%
The partial width of the decay $T\rightarrow B_{c}^{\ast -}B_{c}^{\ast -}$
is determined by the expression%
\begin{equation}
\Gamma \left[ T\rightarrow B_{c}^{\ast -}B_{c}^{\ast -}\right] =\frac{G^{2}%
\widetilde{\lambda }}{2\cdot 80\pi m^{2}}\left( m^{4}-3m^{2}m_{B_{c}^{\ast
}}^{2}+6m_{B_{c}^{\ast }}^{4}\right) ,  \label{eq:PDw2}
\end{equation}%
with $\widetilde{\lambda }$ being equal to $\lambda (m,m_{B_{c}^{\ast
}},m_{B_{c}^{\ast }})$. As a result, we find $\ $%
\begin{equation}
\Gamma \left[ T\rightarrow B_{c}^{\ast -}B_{c}^{\ast -}\right]
=14.0_{-5.3}^{+6.4}~\mathrm{MeV}.  \label{eq:DW2}
\end{equation}

The full width of the tetraquark $T$ saturated by these three decay channels
is
\begin{equation}
\Gamma _{\mathrm{T}}=55.5_{-9.9}^{+10.6}~\mathrm{MeV}.
\end{equation}

\section{Summary}


In present paper we have studied, in a rather detailed form, the tensor
tetraquark $T=bb\overline{c}\overline{c}$ by calculating its mass, current
coupling and full width. The mass of this particle $m=(12.795\pm 0.095)~%
\mathrm{GeV}$ is above a few two $B_{c}^{(\ast )-}B_{c}^{(\ast )-}$ meson
thresholds. As a result, it is unstable against strong decays to $%
B_{c}^{-}B_{c}^{-}$, $B_{c}^{-}B_{c}^{\ast -}$, and $B_{c}^{\ast
-}B_{c}^{\ast -}$ pairs. The full width\ of $T$ \ have been evaluated by
taking into account namely these decay channels.

The dominant decay channel of the tensor tetraquark $T$ is the process $%
T\rightarrow B_{c}^{-}B_{c}^{\ast -}$ with the partial width $(25.2\pm 7.3)~%
\mathrm{MeV}$. The partial widths of another two decays $T\rightarrow
B_{c}^{-}B_{c}^{-}$ and $T\rightarrow B_{c}^{\ast -}B_{c}^{\ast -}$ are
equal to $16.3_{-4.06}^{+4.35}~\mathrm{MeV}$ and $14.0_{-5.3}^{+6.4}~\mathrm{%
MeV}$, respectively. The full width $\Gamma _{\mathrm{T}%
}=55.5_{-9.9}^{+10.6}~\mathrm{MeV}$ of the tetraquark $T$ saturated by these
three channels characterizes it as a four-quark meson with "moderate" width.
This prediction is comparable with our results for the scalar and
axial-vector tetraquarks $bb\overline{c}\overline{c}$. It is smaller than
widths of the scalar tetraquarks, but larger than the width of the
axial-vector state $T_{\mathrm{1}}$. In any case, parameters of the tensor
tetraquark $T$ do not differ considerably from those of scalar and
axial-vector states and fit the class of $bb\overline{c}\overline{c}$ exotic
mesons.

Although spectroscopic parameters of the tetraquarks $bb\overline{c}%
\overline{c}$ were calculated in the context of different models and
approaches, the same cannot be said for their decay channels. But it is
difficult to make credible conclusions about nature of these states without
information on their widths. This is very important, because there are
controversial predictions in the literature concerning stability of the $bb%
\overline{c}\overline{c}$ particles. As is seen, there is plenty to be done
for comprehensive analyses of the four-quark structures $bb\overline{c}%
\overline{c}$ with various spin-parities.

\section*{ACKNOWLEDGEMENTS}

K. Azizi is thankful to Iran National Science Foundation (INSF) for the
partial financial support provided under the elites Grant No. 4025036.

\end{document}